# Enhanced Frequency Conversion in Parity-Time Symmetry Line

Jiankun Hou[1], Jiefu Zhu[2], Ruixin Ma[1], Boyi Xue[1], Yicheng Zhu[1], Jintian Lin[3], Xiaoshun Jiang[4],

Yuanlin Zheng[2], Xianfeng Chen[2], Ya Cheng[4], Li Ge[5]*, and Wenjie Wan[1,2]*

[1]State Key Laboratory of Advanced Optical Communication Systems and Networks, University of Michigan-Shanghai Jiao Tong University Joint Institute, Shanghai Jiao Tong University, Shanghai 200240, China
[2]Department of Physics and Astronomy, Shanghai Jiao Tong University, Shanghai 200240, China
[3]State Key Laboratory of High Field Laser Physics and CAS Center for Excellence in Ultra-Intense Laser Science, Shanghai Institute of Optics and Fine Mechanics, Chinese Academy of Sciences, Shanghai 201800, China
[4]National Laboratory of Solid-State Microstructures, College of Engineering and Applied Science and School of Physics, Nanjing University, Nanjing 210093, China
[5]Department of Physics and Astronomy, College of Staten Island, the City University of New York, NY 10314, USA
*Corresponding authors: Wenjie Wan wenjie.wan@sjtu.edu.cn or Li Ge li.ge@csi.cuny.edu

**Non-Hermitian degeneracies reveal intriguing and non-trivial behaviors in open physical systems. Examples like Parity-Time (PT) symmetry breaking, topological encircling chirality, and enhanced sensing near an exceptional point (EP) are often associated with the abrupt nature of the phase transition around these degeneracies. Here we experimentally observe a cavity-enhanced second-harmonic frequency (SHG) conversion on a PT symmetry line, i.e. a set consisting of open-ended isofrequency or isoloss lines, both terminated at EPs on the Riemann surface in parameter space. The enhancement factor can reach as high as 300, depending on the crossing point whether in the symmetry or the broken phase of the PT line. Moreover, such enhancement of SHG enables sensitive distance sensing with a nanometer resolution. Our works may pave the way for practical applications in sensing, frequency conversion, and coherent wave control.**

Hermiticity of a Hamiltonian is the key postulate of quantum mechanics. It ensures the real eigenvalues and physical observables in ideal closed quantum systems. However, a ubiquitous non-Hermitian nature is generally manifested in most physical systems due to the no-conserving energy or particle flow (gain or loss) with the environment. Among these open systems, a special class of non-Hermitian Hamiltonian with parity-time (PT) symmetry can exhibit entirely real eigenvalues in their symmetric phase. This surprising symmetry has been demonstrated in many physical fields, including optics [1-3], acoustics [4,5], microwaves[6,7], electronics[8-10], opto-mechanics[11-13], and cold atoms[14,15]. Moreover, such PT-symmetric structures have enabled various intriguing features, including unidirectional invisibility [16], pump-induced laser terminations [17], and loss-induced transparency [18]. Essential to these findings is a prominent non-Hermitian degeneracy called exceptional point (EP) [19], which corresponds to the phase transition point at which the eigenvalues of the underlying system and the corresponding eigenvectors simultaneously coalesce. Such EPs representing the exact balance between the internal gain/loss and mode coupling can exhibit square root and other integer-root singularities, making them extremely sensitive to external perturbation [20-22].

Recently, this property of EPs has attracted tremendous interest in sensing scenarios, including single nanoparticle sensing with optical microcavities[22], enhanced Sagnac effect in an optical gyroscope[23,24], and PT-symmetric electromechanical accelerometer[25]. In these works, the attention has been limited to the real part of the energy eigenvalues, i.e. the eigenfrequency splitting $\Delta\omega$, which can be enhanced by orders of magnitude through the $\varepsilon^{1/N}$-dependence (where N is the order of the EP and, $\varepsilon$ is the perturbation)[26-28]. However, few efforts have been made to investigate the drastic change at EP in the imaginary part of the energy eigenvalues, which also plays a crucial role in non-Hermitian lasers [29] and coherent perfect absorbers [30,31]. Moreover, non-Hermitian systems usually are complicated with multiple parameters, forming touching [32,33], intersecting [34], or partially overlapping [35] Riemann sheets in the multiple-parameter space. On these Riemann surfaces, the exact PT symmetry condition often emerges as PT-symmetry lines [36] that feature either isofrequency or isoloss properties but not both [37], in sharp contrast to the EPs. Probing the intrinsic dynamics across these PT-symmetry lines, especially those involving more than one EP, is extremely important for non-Hermitian and topological physics [38-40]. However, the experimental realization and control of such non-Hermitian systems require more degrees of freedom, posing a significant challenge to demonstrating PT-symmetry lines on the Riemann surfaces.

In this work, we theoretically and experimentally demonstrate an enhanced second harmonic generation process in an optical microcavity with second-order nonlinearity, showing that a PT-symmetry line (henceforth referred to as the PT line for short) on the Riemann surface in parameter space can greatly enhance SHG. Moreover, two distinguished scenarios when crossing the PT line are demonstrated for the enhanced SHG, depending on whether the crossing occurs in the symmetric or broken phase of the PT line. The highest SHG enhancement factor is found to be around 300 in the broken phase of the PT line. Moreover, these two crossing scenarios also result in two different mode crossing in the SHG spectra, i.e. normal mode crossing in the broken phase and anti/avoided mode crossing in the symmetry phase, depending on the degeneracy whether existing on the real or imaginary part of the PT line. Based on the enhanced SHG, a sensitive distance sensing scheme with a nanometer resolution has been demonstrated. These results provide a more general paradigm and experimentally feasible scheme to probe the consequences of non-Hermitian degeneracies, opening a new avenue for non-Hermitian physics.

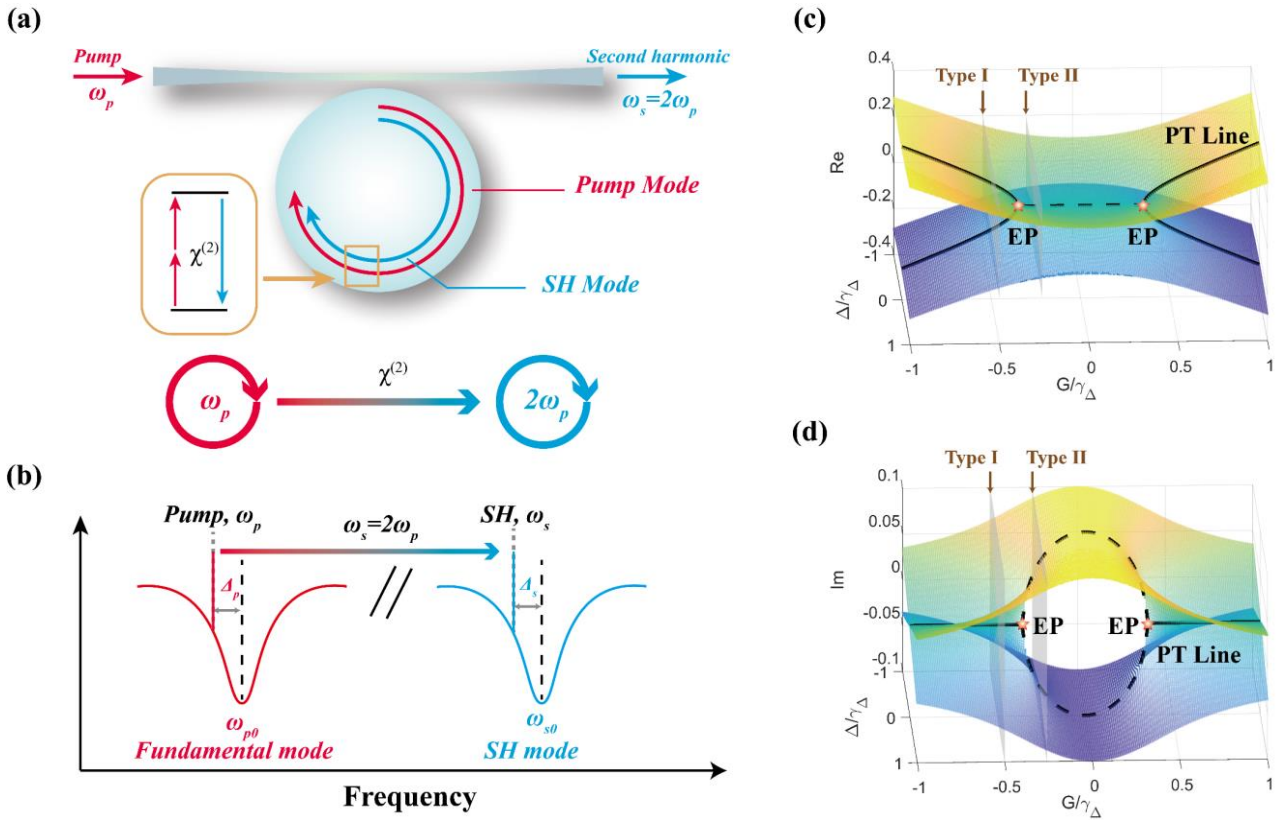

**Figure 1: Parity-Time symmetry line on the Riemann surfaces for the enhanced SHG in an optical microcavity.** (**a**) SHG process in a lithium niobate microcavity with two optical modes: the fundamental mode $\omega_{p0}$ and the SH $\omega_{s0}$ as shown in (**b**). The pump light (red) at $\omega_p$ is converted into its second harmonic light (blue) at $\omega_s$ through the $\chi^{(2)}$ process. The corresponding Riemann surfaces for the real (**c**) and imaginary parts (**d**) of the square-root term of the eigenvalues, as a function of $G/\gamma_\Delta$ and $\Delta/\gamma_\Delta$. The PT symmetry

line, i.e., $\Delta = 0$, is separated into two regimes: the symmetry phase (solid line) and the broken phase (dashed line). There are two types of eigenvalue evolution across the PT symmetry line (grey planes): Type I crossing ($8G^2 - \gamma_\Delta^2 > 0$) intersects the symmetry phase of the PT line, where the imaginary parts of the eigenvalues coincide but the real parts differ at $\Delta/\gamma_\Delta = 0$. Type II crossing ($8G^2 - \gamma_\Delta^2 < 0$) intersects the broken phase, where the real parts of the eigenvalues coincide but the imaginary parts differ at $\Delta/\gamma_\Delta = 0$. Simulation parameters: $\gamma_\Delta = -0.2 GHz$.

The proposed PT symmetry line can be realized in a whispering-gallery-mode (WGM) optical microcavity of quadratic second-order nonlinearity $\chi^{(2)}$ (Fig. 1a), which allows the second harmonic (SH) generation (SHG) from an incoming pump beam with a fundamental wave (FW) frequency $\omega_p$. Inside the microcavity, both the SH and FW waves can individually experience optical resonances, at which their frequency detunings, i.e. $\Delta_s$, $\Delta_p$ from each resonance center are strictly determined by the cavity dispersion and the energy conservation requiring the SH frequency $\omega_s = 2\omega_p$ (Fig. 1b). Effectively, these coupled resonances in the frequency domain can construct a PT symmetry in a synthetic space [13], and these interesting phenomena can be interpreted in a non-Hermitian manner with the help of the PT-symmetry line as follow.

Here the governing Hamiltonian of our system can be formulated as $i\frac{d\psi}{dt} = H\psi$, where $\psi = (\alpha, \beta)^T$ represents the field vectors and $\alpha, \beta$ are the internal cavity amplitudes for FW wave at $\omega_p$, and SH wave at $\omega_s$, respectively. $H$ is a 2 × 2 non-Hermitian Hamiltonian:

$$H = \begin{pmatrix} -\Delta_p - i\gamma_p & -g^*\alpha^* \\ -g\alpha & -\Delta_s - i\gamma_s \end{pmatrix} \quad (1)$$

where the subscript "p" and "s" represent the pump and the SH, respectively. $\gamma_{p,s}$ are their effective cavity decay rates due to internal absorption, scattering, radiation loss, and external coupling, which are typically different for the two wavelengths; $g$ is the second-order nonlinear coupling constant determined by $\chi^{(2)}$, mode volume and field overlap. $\Delta_{p,s} = \omega_{p,s} - \omega_{p0,s0}$ are the frequency detunings from their neighboring WGMs, which are related by $\Delta_s - \Delta_p = \Delta_p - \delta$ and $\delta = \omega_{s0} - 2\omega_{p0}$ is the frequency mismatch between two cavity modes. $\Delta_p$ is controlled by scanning the pump frequency, while $\delta$ can be tuned by changing the photorefractive effect via the pump power [ details in the supplement Ref. 43]. The supermodes of the Hamiltonian are hybridization between the fundamental and its second harmonic. Note that, $H$ is a *nonlinear* Hamiltonian as well, whose off-diagonal terms, e.g. $-g^*\alpha^*$, $-g\alpha$, depend on the internal fields $\alpha$ and the second-order nonlinear coupling constant $g$.

Under linear approximation, solving the Hamiltonian gives two eigenvalues:

$$\begin{aligned}\lambda_\pm &= \frac{1}{2}\left[\left(-\Delta_p - \Delta_s - i\gamma_p - i\gamma_s\right) \pm \sqrt{\left(\Delta + i\gamma_\Delta\right)^2 + 4G^2}\right] \\ &= \frac{1}{2}\left[\left(-\Delta_p - \Delta_s - i\gamma_p - i\gamma_s\right) \pm \sqrt{\left(\Delta_p - \delta + i\gamma_\Delta\right)^2 + 4G^2}\right]\end{aligned} \quad (2)$$

where $\Delta = \Delta_s - \Delta_p$, $\gamma_\Delta = \gamma_s - \gamma_p$, $G = |g\alpha|$. Obviously, the dependence of the eigenfrequency on the system parameters is determined by the square-root term $\eta_\pm = \pm\frac{1}{2}\sqrt{\left(\Delta + i\gamma_\Delta\right)^2 + 4G^2}$ in Eq. (2). For simplicity, we only consider the square-root term, and Fig.1c & 1d show the complex Riemann surfaces of $\eta_\pm$ over the ($\Delta/\gamma_\Delta$, $G/\gamma_\Delta$)-plane. If the detuning term vanishes, i.e. $\Delta = 0$, this Hamiltonian returns to the well-known PT symmetry condition, represented by singularity lines in the Riemann surface, termed as PT symmetry line [36]. As usual, the PT symmetry line can be divided into the PT-symmetric phase ($4G^2 - \gamma_\Delta^2 > 0$ solid line) and the broken phase ($4G^2 - \gamma_\Delta^2 < 0$ dash line), separated by an exceptional point at $4G^2 = \gamma_\Delta^2$. Note that, the linear approximation is valid near the PT line, for larger pumps or detunings, the nonlinearity should be considered [40].

More interestingly, for a constant $G$, solely tuning the detuning term $\Delta$ can result in two distinct dynamics, i.e. type I and type II crossings, depending on the intersection with the symmetric or broken phase of the PT line[37]: for the type I crossing as shown in Fig. 2a, when $4G^2 - \gamma_\Delta^2 > 0$, two eigenvalues on the Riemann surfaces only intersect in the imaginary part, right at the symmetric phase of the PT line. In contrast, the type II crossing ($4G^2 - \gamma_\Delta^2 < 0$), instead, only permits such intersection occurring in the real part (Fig. 2b). In Eq.(2), if $\gamma_\Delta = 0$, the eigenvalues $\eta_\pm = \pm \frac{1}{2}\sqrt{\Delta^2 + 4G^2}$ can form a double cone topology with a linear dependence of $\Delta$ and $G$ and this creates a diabolic point (DP) at $\Delta = G = 0$. On the other hand, for the case $\Delta = 0$, the eigenvalues $\eta_\pm = \pm \frac{1}{2}\sqrt{4G^2 - \gamma_\Delta^2}$ show the PT symmetry line and create an EP at $4G^2 = \gamma_\Delta^2$, where both the real and the imaginary parts of the eigenvalues degenerate and $\eta_\pm$ exhibits a square-root parameter dependence. It is well-known that a double-resonant scenario, i.e. $\Delta_s = \Delta_p = 0$ can lead to tremendous SHG enhancement in a conventional optical cavity[41,42]. As shown in Fig. 2d, the double resonance can enhance SHG up to ~100 as compared to the non-resonant case. However, it is often underappreciated that such a double resonant scheme can instead result in a surprising mode splitting of SHG (Fig. 2c) due to a strong FW pumping or reduced loss in the cavity.

Previously, many prior works emphasized the EP enhancement for various sensing themes as compared to the DP case [44]. Here similar trends can be expected in our case of crossing PT lines for the SHG. In the case of crossing the PT line near $\Delta = 0$, the eigenvalues read $\eta_\pm \approx \pm \frac{1}{2}\sqrt{i2\Delta\gamma_\Delta + 4G^2 - \gamma_\Delta^2}$. If $|4G^2 - \gamma_\Delta^2|$ is small, i.e. near the EP, $\eta_\pm$ are mainly dependent on $\Delta$ or $\gamma_\Delta$ (which can be affected by external perturbations such as nanoparticles [22] ), exhibiting a square-root behavior similar to the EP case[21]. However, when $|4G^2 - \gamma_\Delta^2|$ becomes large, enough, the square-root dependence of $\gamma_\Delta$ turns weak. Therefore, the maximum sensitivity or enhancement can be found near the EP and decrease as the intersection point moves away from the EP. For example, in the type I crossing shown in Fig. 2a, $|4G^2 - \gamma_\Delta^2| = 0.02 GHz$, the evolution of eigenvalues is relatively smooth at the intersection. On the contrary, at the type II crossing shown in Fig. 2b, $|4G^2 - \gamma_\Delta^2| = 0.0008 GHz$, the evolution of eigenvalues becomes much steeper at the intersection. In this case, a sharp SHG enhancement can be expected.

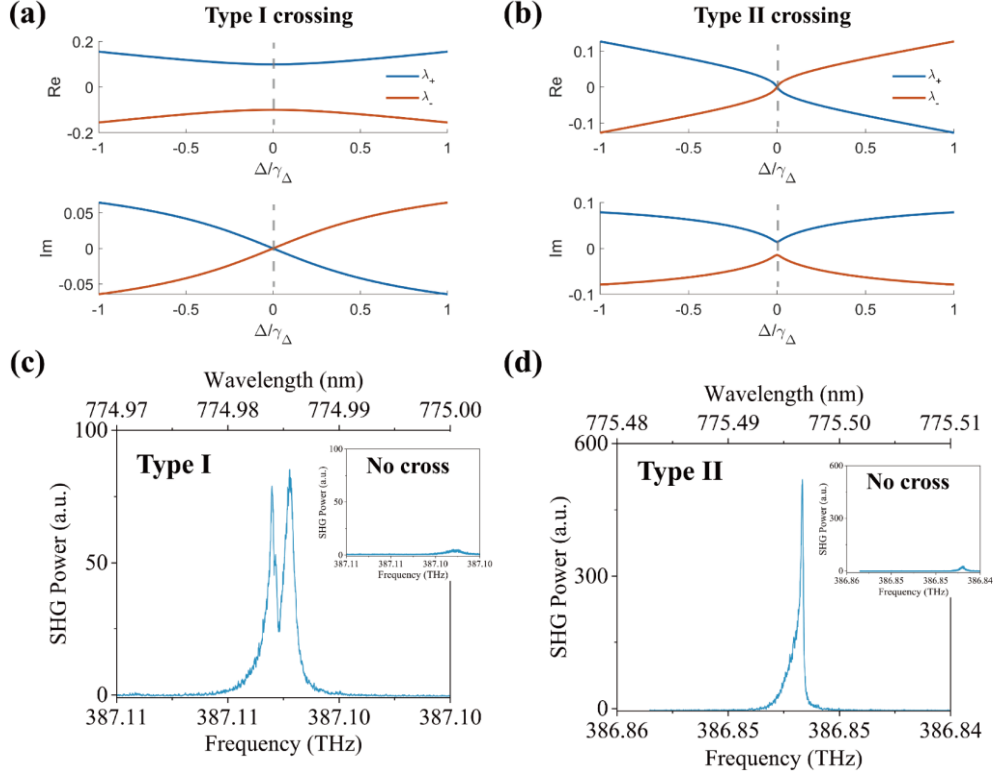

**Figure 2: Two crossing scenarios of the PT Line.** There are two types of eigenvalue evolution across the PT symmetry line (grey planes): **(a)** Type I crossing ( $4G^2 - \gamma_\Delta^2 > 0$ ) intersects the symmetry phase of the PT line, where the imaginary parts of the eigenvalues coincide but the real parts differ at $\Delta/\gamma_\Delta = 0$, leading to avoided crossing in SHG spectra in **(c)**. **(b)** Type II crossing ( $4G^2 - \gamma_\Delta^2 < 0$ ) intersects the broken phase, where the real parts of the eigenvalues coincide but the imaginary parts differ at $\Delta/\gamma_\Delta = 0$, leading to the mode superposition in SHG spectra in **(d)**. Corresponding experimental results are shown in **(c)** and **(d)**. Upper insets: Far away from the crossing ($\omega_{s0} < 2\omega_{p0}$) , SHG mainly from single resonance enhancement. Simulation parameters: $\gamma_\Delta = -0.2 GHz$ .

In our experiment, a z-cut thin-film lithium niobate microdisk with ~100μm diameter and Q-factor ~$10^6$ is fabricated by photolithography and chemo-mechanical polishing (CMP) [45,46]. The FW signal around the 1550nm telecommunication band is coupled into the microcavity through a tapered fiber. On the output port, the FW and SHG are separated by a 780/1550 nm wavelength division multiplexer (WDM), and separately received by an InGaAs photodetector (PD) and a photomultiplier tube (PMT). The tapered fiber is precisely positioned on a piezoelectric stage (~10 nanometers resolution) to tune the gap (~100nm) from the cavity. Different coupling conditions can be controlled by the position of the contact point on the edge of the microdisk, resulting in different coupling losses to control $\gamma_\Delta$ term [43]. Here we utilize an inherent photorefractive effect of lithium niobate materials [47,48] to tune the detuning term $\delta$ in Eq. (2), a crucial technique in our experiment (details in the supplement). The scanning frequency of the pump laser is ~20Hz, much faster than the rate of photorefractive decay [47,48]. Therefore, the photorefraction can be assumed to be proportional to the input FW's power, which can be controlled by a variable optical attenuator (VOA).

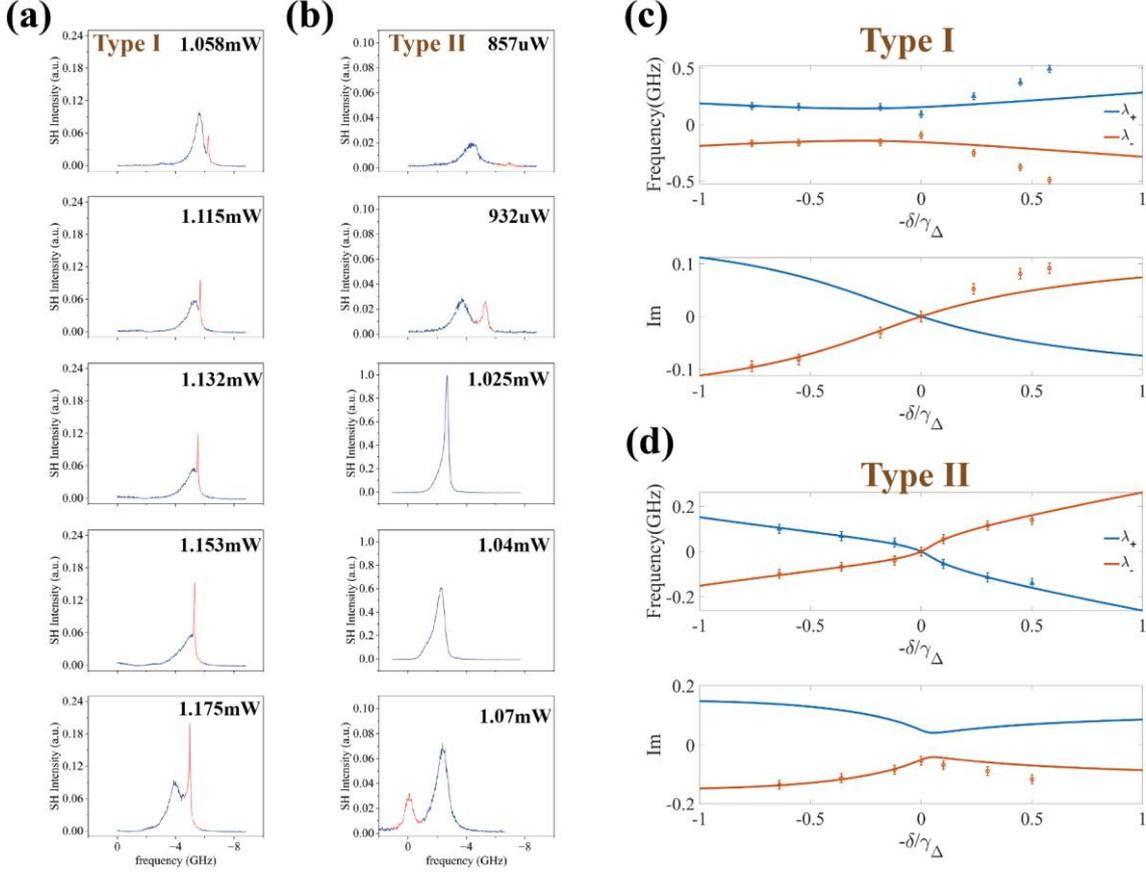

**Figure 3: Experimental observation of the type I and type II crossings of the PT symmetry line.** Experimental results of SHG spectra at **(a)** the type I and **(b)** type II crossings. The small pump power difference can lead to a large disturbance to the detuning difference via thermo-optical photorefractive effects. The eigenmodes "-" (red peak) and "+" (blue peak) show anti-/avoided mode crossing at the type I scenario **(a)**, but cross with each other at the type II crossing **(b)**. **(c)** and **(d)** The real parts and imaginary parts of the square-root term of the eigenfrequencies measured through the SHG spectra near the PT line, i.e. $\delta \approx 0$. The top parts of **(c)** depict the real parts of the avoided mode crossing, while the "-" mode in the imaginary parts first surges rapidly near the AVS point and saturates afterward. **(d)** exhibits a clear mode crossover and a clear SHG peak at the PT line. Error bars represent the uncertainty in the measurement due to the system jitter. Circles and triangles: experimental results. Solid lines are theoretically calculated through coupled mode theory. Numerical parameters: $\gamma_p = 0.45 GHz$, $\gamma_s = 0.2 GHz$ in **(c)**; $\gamma_p = 0.725 GHz$, $\gamma_s = 0.425 GHz$ in **(d)**.

In both crossing scenarios, the SHG is mainly generated through cavity resonance enhancement. When scanning the FW laser wavelength, two SHG peaks appear: one near the FW resonance mode ($\omega_{p0}$) and the other near the SH mode ($\omega_{s0}$). In the case of large detuning, i.e. $\Delta \gg \gamma_\Delta, G$, the real part of eigenvalue in Eq.(2) reveals two solutions for the above two cases, i.e. $\omega_p = \omega_{p0}$ for $\lambda_+$ and $\omega_s = \omega_{s0}$ for $\lambda_-$. In this case, either the FW mode or the SHG mode is on resonance only, but not both. This singlet resonance scenario can be clearly observed in the SHG spectra in Fig. 3a during the FW laser scanning, where the "+" mode exhibits a slightly higher SHG than the "-" case. In order to obtain a higher conversion efficiency, it is natural to merge the two resonance modes together. Effectively, this requires minimizing the square-root term in Eq. (2), so that when the FW laser is on resonance ($\Delta_p = 0$), the mode splitting between the "+" and "-" SHG modes can be reduced by minimizing $\delta$. Experimentally, $\delta$ can be tuned by precisely varying the input pump's power through photorefractive effects as mentioned[48]. Here the FW and SHG modes react to the input pump differently and $\delta$ can be assumed to be almost linearly related to the pump power calibrated in the supplement [43]. Note that, the loss term $\gamma_\Delta$ remains constant during the tuning process.

Figure 3 shows the SHG spectra for both type I (Fig. 3a,c) and type II (Fig. 3b,d ) crossing scenarios at different pump powers. Here the variation of the pump remains relatively small within 10%, but their induced changes in $\delta$ are much larger. For type II crossing (Fig. 3b), initially, the "+" mode (blue) and the "-" mode are well separated along the frequency axis due to $\omega_{s0} < 2\omega_{p0}$. To facilitate analysis, we set the scanning start frequency of the laser as the frequency zero of the horizontal axis. When increasing the pump power, effectively reducing $\delta$, $\omega_{s0}$ approaches $2\omega_{p0}$. Two split SHG resonance peaks merge into one single peak near the condition $P_{in} \approx 1.025 mW$, where the intensity of the "-" mode suddenly surges to a maximum SHG enhancement of over 300 times as compared to the pre-merge case. However, further increase of the pump leads to the walk-off of the two modes again.

It is *not* a straightforward task to switch to the type I crossing by simply increasing $G$ to reach the condition of $4G^2 - \gamma_\Delta^2 > 0$, since the detuning term $\delta$ also depends on the input pump's power. Experimentally, we find it more feasible to reduce $\gamma_\Delta$ instead. The loss terms $\gamma_p$ and $\gamma_s$ can be controlled by moving the contact point between the tapered fiber and the microcavity (supplemental Ref. 40). In this manner, we have been able to minimize $\gamma_\Delta$ to around $-0.25 GHz$. Like the type II crossing, the initial two split modes in the SHG spectra are getting closer to each other when approaching the PT line, and the "-" mode is amplified during the process (Fig. 3a). However, in sharp contrast to the type II crossing, the two modes tend not to cross with each other at the PT line, exhibiting an avoided/anti mode crossing (AMC) in Fig. 3c [37]. More interestingly, the intensity peak of the "-" mode continues to surge even after passing the AMC point and finally saturates, this is also consistent with the calculated results of the imaginary part as shown in Fig. 3c. In comparison, the SHG only peak at the crossing point in the type II crossing (Fig. 3d). Here the overall SHG enhancement factor is only 5, not as large as the previous type II crossing case since the average imaginary part of type I crossing is below the type II case's as shown in Fig. 1d.

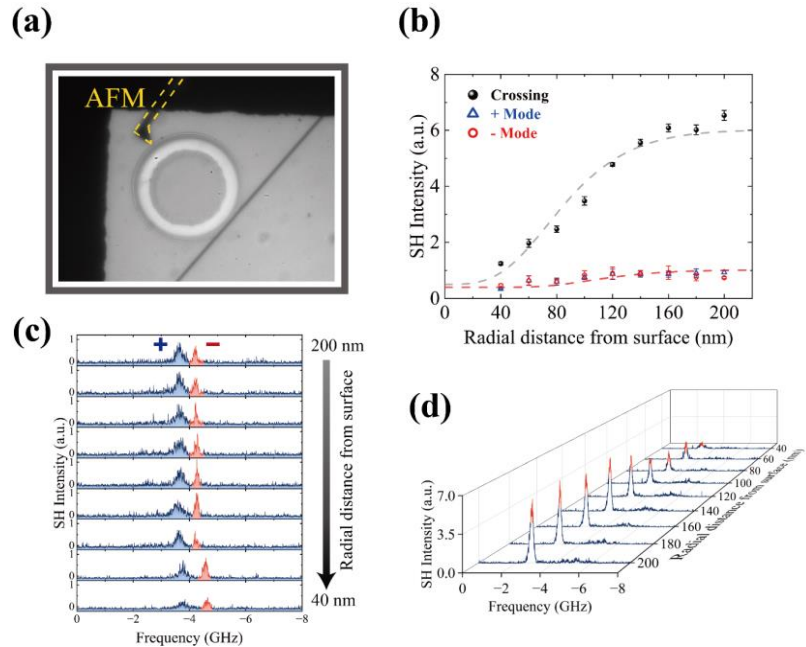

**Figure 4: PT-line enhanced distance sensing. (a)** Optical microscope image of the LN microcavity and the Au-coated AFM. **(b)** Experimental results of SH intensity dependence of the tip's radial position. The results are compared with the PT-line crossing case (black) and the one away from the PT line (blue and red) in the type II crossing, and their corresponding spectra are depicted in **(c)** and **(d)**. Dashed lines are fitted curves. Pump power: 1 mW. Error bars denote the standard deviations of three measurements.

Like the EP [49,50], the PT line could also potentially enhance the sensing applications. As a proof-of-principle experiment (Fig. 4), we demonstrate a PT-line enhanced nanometer-resolution distance sensing near the PT line of a type II crossing. An Atomic-Force-Microscope (AFM) tip mounted on a nano-translation stage is externally approaching the microcavity to perturb the SHG (Fig. 4a). Both loss factors and frequency detuning mismatch are affected as the tip approaches. The system initially operates at the PT line and rapidly decays its SH intensity as the tip approaches (Fig. 4b,d). In this manner, the gap distance can be precisely measured with the SHG, and the corresponding resolution can be obtained around 0.4nm (details in the supplement). In comparison, for an off-PT line case (Fig. 4b,c), both "-" and "+" modes demonstrate only marginal changes influenced by external perturbations, either in the amplitude (Fig. 4b) or the mode splitting (Fig. 4c). From the fitted curve, it can be inferred that the relationship between eigenvalues and $\gamma_\Delta$ is approximately $\gamma_\Delta^{0.7}$-dependence near the PT line and almost liner dependence in off-PT line case.

In conclusion, we have experimentally demonstrated a cavity-enhanced second-harmonic frequency conversion in a PT symmetry line, where the two crossing scenarios reveal dramatic distinct crossing dynamics and enhancement factors. As compared to prior works of EPs, the PT line is much easier to access experimentally than the singularity condition such as the EP. In the future, extra assistance such as nanoparticles may help to control the loss term $\gamma_\Delta$, so that the enhancement factor can be further improved. Our work provides a new perspective on the enhancement of frequency conversion in non-Hermitian physics, paving the way for practical applications in sensing, frequency conversion, and coherent wave control.

**Acknowledgment**: This work was supported by the National Science Foundation of China (Grant No. 12274295, No. 92050113); the National Key R&D Program of China (Grant No. 2023YFB3906400, No. 2023YFA1407200).